\documentclass[11pt]{article}

\usepackage{amsfonts,epsfig,fullpage,times}

\newtheorem{theo}{Theorem}

\newtheorem{lem}{Lemma}
\newtheorem{claim}[lem]{Claim}
\newtheorem{coro}[lem]{Corollary}

\newtheorem{define}{Definition}

\newtheorem{alg}{Algorithm}

\newcommand{\BE}{\begin{enumerate}} \newcommand{\EE}{\end{enumerate}}
\newcommand{\BI}{\begin{itemize}} \newcommand{\EI}{\end{itemize}}
\newcommand{\BDes}{\begin{description}}\newcommand{\EDes}{\end{description}}
\newcommand{\BT}{\begin{theo}} \newcommand{\ET}{\end{theo}}
\newcommand{\BL}{\begin{lem}} \newcommand{\EL}{\end{lem}}
\newcommand{\BD}{\begin{define}} \newcommand{\ED}{\end{define}}
\newcommand{\BCM}{\begin{claim}} \newcommand{\ECM}{\end{claim}}
\newcommand{\BC}{\begin{coro}} \newcommand{\EC}{\end{coro}}
\newcommand{\BA}{\begin{alg}} \newcommand{\EA}{\end{alg}}

\def\FullBox{\hbox{\vrule width 8pt height 8pt depth 0pt}}
\newcommand{\qed}{\;\;\;\FullBox}
\newenvironment{proof}{\noindent{\bf Proof:~~}}{\(\qed\)}
\newcommand{\BPF}{\begin{proof}} \newcommand {\EPF}{\end{proof}}
\newenvironment{proofof}[1]{\noindent{\bf Proof of {#1}.~}}{\endproof}
\newcommand{\BPFOF}{\begin{proofof}} \newcommand {\EPFOF}{\end{proofof}}

\newcommand{\BEQN}{\begin{eqnarray}}\newcommand{\EEQN}{\end{eqnarray}}
\newcommand{\BEQ}{\begin{equation}} \newcommand{\EEQ}{\end{equation}}

\renewcommand{\Pr}{{\rm Pr}}

\newcommand{\eps}{\epsilon}

\newcommand{\remove}[1]{}
\newcommand{\ignore}[1]{}

\usepackage{fancyhdr,url}
\usepackage{amssymb}

\newcommand{\eat}[1]{}

\def\blackslug{\hbox{\hskip 1pt \vrule width 4pt height 8pt
     depth 1.5pt \hskip 1pt}}
\def\QED{\quad\blackslug\lower 8.5pt\null\par}

\def\Proof{\par\noindent{\it Proof:~}}
\def\proof{\Proof}

\newcommand{\maxindex}{\mbox{\rm max-index}}

\title{Bounding the Bias of Tree-Like Sampling in IP Topologies}

\author{
    Reuven Cohen
    \thanks{
        Reuven Cohen is with the Deparment of Electrical
        and Computer Engineering, Boston University, Boston, MA
        02215, USA.
    {\tt cohenr@shoshi.ph.biu.ac.il}}
   \hspace*{1cm}
    Mira Gonen\thanks{
        Mira Gonen is with the School of Electrical Engineering,
    Tel Aviv University,
         Ramat Aviv 69978, Israel.
    {\tt gonenmir@tau.ac.il}}
      \hspace*{1cm}   
    Avishai Wool\thanks{
         Avishai Wool is with the School of Electrical Engineering ,
    Tel Aviv University,
         Ramat Aviv 69978, Israel.
    {\tt yash@acm.org}}
}

\date{\today}

\begin{document}
\maketitle

\begin{abstract}
  It is widely believed that the Internet's AS-graph degree
  distribution obeys a power-law form. Most of the evidence showing
  the power-law distribution is based on BGP data.  However, it was
  recently argued that since BGP collects data in a tree-like fashion,
  it only produces a sample of the degree distribution, and this sample
  may be biased. This argument was backed by simulation data and
  mathematical analysis, which demonstrated that under certain
  conditions a tree sampling procedure can produce an artificial
  power-law in the degree distribution. Thus, although the {\em observed\/}
  degree distribution of the AS-graph follows a power-law, this
  phenomenon may be an {\em artifact\/} of the sampling process.

  In this work we provide some evidence to the contrary. We show, by
  analysis and simulation, that when the underlying graph degree
  distribution obeys a power-law with an exponent $\gamma>2$, a
  tree-like sampling process produces a negligible bias in the sampled
  degree distribution. Furthermore, recent data collected from the
  DIMES project, which is not based on BGP sampling, indicates that the
  underlying AS-graph indeed obeys a power-law degree distribution
  with an exponent $\gamma>2$. By combining this empirical data with our
  analysis, we conclude that the bias in the degree distribution
  calculated from BGP data is negligible.
\end{abstract}

{\bf Classification:} Network Computing, Internet Topology Models



\section{Introduction}
\subsection{Background and Motivation}
The connectivity of the Internet crucially depends on the
relationships between thousands of Autonomous Systems (ASes) that
exchange routing information using the Border Gateway Protocol (BGP).
These relationships can be modeled as a graph, called the AS-graph,
in which the vertices model the ASes, and the edges model the
peering arrangements between the ASes.

\eat{
Besides its inherent interest, modeling the AS-graph is an important
{\em practical} goal. The reason is that with more accurate topology
models, we can build more accurate synthetic network topology
generators. Topology generators are widely used whenever one wishes
to evaluate any type of Internet-wide phenomenon that depends on BGP
routing policies, namely, traceroute-based sampling. A few recent
examples include testing the survivability of the Internet
\cite{ajb00,djms03}, comparing methods of defense against Denial of
Service (DoS) attacks \cite{wlc05}, and suggesting new methods for
combating source IP address spoofing \cite{lps04}.
}

Significant progress has been made in the study of the AS-graph's
topology over the last few years. A great deal of effort has been
spent measuring topological features of the Internet. Numerous
research projects~\cite{fff99,ab00,ccgjsw02,lbcx03,wgjps02,wj02,bt02,
  bgw05b,bgw04,krr01,ss05,lc03,bb01,bbbc01,rn04,tpsf01,tgjsw02,gt00,bs02}
have ventured to capture the Internet's topology. Based on these and
other topological studies, it is widely believed that the Internet's
degree distribution has a power-law form with an exponent
$2<\gamma<3$, i.e., the fraction of vertices with degree~$k$ is
proportional to $k^{-\gamma}$.  Most of these studies are based
on BGP data from sources such as RouteViews \cite{routeViews} and CIDR
Reports \cite{CIDR}.

\eat{
collection
procedure.  Moreover, most studies infer the Internet's topology from
the union of traceroutes from a single root computer to a larger
number of (or all) other computers in the network.
}

\subsection{Related Work}
It was recently argued~\cite{lbcx03,cm04,cm04b,pr04,ackm05,ss05,
  ccgjsw02,tgjsw02,wgjps02} that the evidence obtained from the
analysis of the Internet graphs sampled as described above may be
biased.  In a thought-provoking article, Lakhina et al.\ \cite{lbcx03}
claimed that a power-law degree distribution may be an artifact of the
BGP data collection procedure.  They suggest that although the {\em
  observed\/} degree distribution of the AS-graph follows a power-law
distribution, the degree distribution of the real AS-graph might be
completely different. They claim that with tree-like sampling, such as
that employed by BGP, an edge is much more likely to be visible, i.e.,
included in the sampled graph, if it is close to the root.  Moreover,
in tree-like sampling, high-degree vertices are more likely to be
encountered early on, and therefore they are sampled more accurately
than low-degree vertices. They backed this argument with simulations
that indicated that under some conditions, a BFS sampling process {\em
  in itself} is sufficient to produce a power-law degree distribution
in the sample, even when the underlying graph is a sufficiently-dense
Erd\H{o}s-Renyi \cite{er60} graph.

Subsequently \cite{cm04,cm04b} gave a mathematical foundation to the
argument of \cite{lbcx03}. They showed that BFS tree sampling produces
a power-law degree distribution, with an exponent of $\gamma=1$, both
for Poisson-distributed random graphs and for
$\delta$-regular random graphs. In other words, they showed that a
tree sampling process may have a significant bias, and may produce an
artificial power-law---albeit with an exponent that is very different
from that observed in the AS-graph.

On the other hand, Petermann and De Los Rios \cite{pr04} showed that
for single-source tree sampling of a BA graph \cite{ba99}, the
exponent obtained for the power-law distribution is only slightly
under-estimated. This cannot be viewed as strong evidence against the
argument of \cite{lbcx03}, since the analysis assumes a BA-model,
which is a highly idealized evolution model of the AS-graph. However,
this result does indicate that at least in some power-law graphs, a
tree sample does not create a significant bias in the degree distribution.

More recently, \cite{ackm05} analyzed the degree distribution
discovered by a BFS tree sampling process over a general graph. Among
other results, they gave a general, but rather unwieldy, expression of
the degree distribution of the sampled graph, depending on the
underlying graph degree distribution. This work is the starting
point of our analysis: we use the results of \cite{ackm05} to
analyze the sampled degree distribution when the underlying graph has
a power-law degree distribution.

A new development in the empirical measurement of the Internet
topology was suggested recently by Shavitt and Shir \cite{ss05}.  In
this work they describe an Internet mapping system called DIMES. DIMES
is a distributed measurement infrastructure for the Internet that is
based on the deployment of thousands of light weight measurement
agents around the globe. Unlike BGP data, that is sampled in a
tree-like fashion, DIMES executes {\tt traceroute}s among all pairs of
its agents, collects the router-level results, and aggregates the
AS-graph from this data. Because of this measurement methodology,
DIMES discovers significantly more links than BGP-based systems.
The salient point for our purposes is that DIMES data too shows a
power-law degree distribution, with an exponent $2<\gamma<3$ ---
and since the DIMES system uses an all-pairs measurement paradigm,
it is difficult to claim that the power-law is an artifact of a
tree-like sampling.

\subsection{Contributions}

Our main contribution is our analysis of the degree distribution
observed in the BFS tree sample, when the underlying graph has a
power-law distribution with an exponent $2<\gamma<3$. Under these
conditions, we prove that the bias in the power-law is negligible:
with high probability, the degree distribution of the high-degree
nodes in the sample also exhibits a power-law, with exactly the same
exponent $\gamma$. We validate our mathematical analysis with
simulation results using the DIMES-measured AS-graph as the underlying
power-law graph.

Putting this result in the context of the Internet topology, we recall
the data collected from the DIMES project is not based on BGP-style
tree sampling. Nevertheless, DIMES data indicates that the underlying
AS-graph indeed obeys a power-law degree distribution with an exponent
$\gamma>2$. By combining this empirical data with our analysis, we
conclude that the bias in the degree distribution calculated from BGP
data is negligible.

\emph{Organization:} In the next section we give an overview of the
results of \cite{ackm05} we rely on. In Section~\ref{sec:DegDist} we
show our main result, that the bias in the degree distribution of a
tree-sampled power-law graph is negligible. In Section~\ref{sec:rig}
we sketch an alternative, more rigorous, analysis of a weaker result,
that validates some of the approximations we used in our main result.
Section~\ref{sec:Implementation} describes the results of our
simulations. We conclude with Section~\ref{sec:Conclusions}.

\section{Highlights of \cite{ackm05}}
\subsection{The General Framework}
The proof of our result is based on the model, sampling process, and
the main results described in \cite{ackm05}. In this section we give a
brief introduction to the main results we need.

\noindent{\bf Notation} Throughout the paper we use $G=(V,E)$ to denote the
underlying graph, and $n=|V|$ to denote the number vertices.
\begin{define}
We say that $\{a_j\}$ is a degree distribution of $G$ if $G$ contains
$a_j\cdot n$ nodes of degree $j$.
\end{define}

In the \cite{ackm05} model, the graph $G$ is not a given graph but
a random graph chosen out of a family of graphs obeying a given degree
distribution $\{a_j\}$. The basic setting is the
configuration model of \cite{b85}: for each vertex of degree $k$ we
create $k$ copies, and then define the edges of the graph according
to a uniformly random matching on these copies.

\subsection{The BFS Tree Sampling Process} \label{sec:ackmBFS}

The \cite{ackm05} model defines a randomized process, that
simultaneously produces a random graph $G$ obeying the degree
distribution $\{a_j\}$, and a BFS tree $T$ that represents the sample.
Note that for a given graph $G$, a BFS algorithm is a deterministic
algorithm, but different outcomes are possible, depending on the order
in which outgoing edges are traversed. In \cite{ackm05} model, a
random choice determines this order.

The sampling process is thought of taking place in continuous time.
However, for technical reasons the authors define a non-standard
notion of time, which we denote by a capitalized word (Time). In this
model, the BFS sample process starts at Time $t=1$ with an empty tree
$T$. As the sampling process evolves, Time decreases to $t=0$, when
the sample tree $T$ includes all $n$ nodes (assuming $G$ is connected).

Before the process starts, for each vertex $v$ there are $deg(v)$
{\em copies} of $v$. Each copy is given a real-valued {\em index} chosen
uniformly at random from the unit interval $[0,1]$. Namely, vertex
$v$ has $deg(v)$ indices, chosen uniformly and independently at
random from the unit interval $[0,1]$. At every Time step $t$ two
copies are matched: one copy is a copy of a vertex already
discovered, and the other copy is a copy with index $t$. Such a
matched pair forms an edge of the original graph. According
to~\cite{ackm05} at Time $t$ the indices of the unmatched copies are
uniformly random in $[0, t)$. Let the {\em maximum index} of a vertex be
the maximum of all its copies' indices. Then at any Time $t$, the
vertices that have not been discovered yet are precisely those whose
maximum index is less than $t$. An edge will be {\em visible},
namely, included in the BFS tree, if at the Time its endpoints are
matched one of them is a copy of an undiscovered vertex.

 \subsection{Useful Notations and Theorems of \cite{ackm05}}
Let $v_t$ be the vertex that has a copy with maximum index $t$.
 Denote by $P_{vis}(t)$ the probability that {\em another} edge outgoing
 from $v_t$ appears in the BFS tree. Namely, $v_t$ is the vertex
 that was discovered at Time $t$, and we are interested in the
 probability that {\em another} vertex is discovered through $v_t$.

 Using the expectation of $P_{vis}(t)$ as approximation we get from
 Equation (7) and Lemma 3 of \cite{ackm05} the following Theorem:
 \footnote{In Lemma 3 of \cite{ackm05} there is a requirement that
  $a_j=0$ for $j<3$. This implies that with high probability the graph is connected.
   However, if we assume that the graph is connected, which is the case we are interested in,
   this requirement can be relaxed. }
 \begin{theo} \label{thm:ackm}
 Let $G$ be a connected graph and let $\{a_j\}$ be a degree
 distribution that is upper bounded by a power-law with an exponent
 larger than 2. Let $\mu=\sum_j{ja_j}$ be the mean degree of $G$. Then it holds that \begin{eqnarray}\label{eq:pvis-approx}
 P_{vis}(t) &\approx&
 \frac{1}{\sum_j{ja_jt^j}}\sum_k{ka_kt^k\left(\frac{\sum_j{ja_jt^j}}{\mu
 t^2}\right)^k}.
\end{eqnarray}
 \end{theo}

\section{The Degree Distribution of the Sampled Graph} \label{sec:DegDist}

Our goal in this section is to show that the bias observed in BFS tree
sampling regarding the degree distribution of the sampled graph is not
significant, when the underlying graph degree distribution obeys a
power-law with an exponent $\gamma>2$. We show the above by examining
the BFS tree received by the sampling process of \cite{ackm05},
described in the previous section. Finding a BFS tree from a single
source is an idealization of the BGP data collection process.  Thus,
if the bias is negligible when using such BFS process, then we argue
that it is very likely to be negligible when using the more general case of
BGP.

Recall that we focus on a BFS process on a random graph $G$.
Let $T$ be the BFS tree received by this process.  Let $deg_T(v)$
denote the degree of node $v$ in the BFS tree received from the
sampled graph. Let $deg_G(v)$ denote the degree of node $v$ in the
graph.

\begin{define}
We say that a vertex $v$ has a {\em high-degree} if $deg_G(v)\ge 18$.
We say that an event occurs {\em with high probability} (w.h.p) if it occurs
with probability at least 0.16.
\end{define}

\noindent The main theorem we prove is the following:
\begin{theo}\label{thm:power-law}
If the underlying graph degree distribution obeys a power-law with
an exponent $\gamma$, where $2<\gamma<3$, then with w.h.p the degree
distribution of the high-degree vertices of the sampled graph also
follows a power-law, with the same exponent value.
\end{theo}
\eat{\begin{theo}\label{thm:degT} Let $m=\frac{i-1}{2(1+d(d-2))}$,
where $d$ is a free parameter that is larger than 2. Then for every
node $v$ it holds that
\[\Pr\left[deg_T(v)\ge m \mid deg_G(v)=i\right]\ge 1-\epsilon,\]
where $\epsilon = e^{-\frac{i-1}{8(1+d(d-2))}}$.
\end{theo}}

We prove Theorem~\ref{thm:power-law} using several Lemmas and
Theorems. Throughout the remainder of this paper we use the following
setting:

\begin{define}\label{def:setup}
  Let $a_k = C\cdot k^{-\gamma}$ be a degree distribution, where
  $2<\gamma<3$ and $C>0$ is a constant normalization factor.  Let
  $\mu$ denote the mean graph degree, i.e., $\mu=E[deg_G(v)]
  =\sum_j{ja_j}$.
\end{define}
Note that since the degree distribution is a power-law with an
exponent larger than 2, in this case $\mu$ is finite.

\noindent Our starting point is Theorem~\ref{thm:ackm} \cite{ackm05}.
Our first step is to approximate the following sum, which
appears in Equation~(\ref{eq:pvis-approx}), when the degree
distribution obeys the power-law of Definition~\ref{def:setup}:
\begin{equation}
  \label{eq:basic_sum}
  \sum_k{ka_kt^k} = C\sum_k{kk^{-\gamma}t^k} = C\sum_k{k^{1-\gamma}}t^k.
\end{equation}

\begin{lem}\label{lem:int}
 Let $\mu$, $a_k$, and $\gamma$ be as in Definition~\ref{def:setup}.
Then
\[
\sum_k{k^{1-\gamma}}t^k\approx \frac{t^3}{\gamma-2},
\] for all $0\le t \le 1$.
\end{lem}
\proof
\[\sum_k{k^{1-\gamma}t^k}\approx \int_1^\infty{x^{1-\gamma}t^x}dx\]
Let $g(t) = \int_1^\infty{x^{1-\gamma}t^x}dx$. Then for all $i\ge 1$
we have that the $i$th derivative of $g(t)$ is
\[
g^{(i)}(t) =
\int_1^\infty{x^{1-\gamma}\prod_{j=0}^{i-1}{(x-j)}t^{x-i}}dx.
\]
Let us now evaluate $g$, and its derivatives, at the boundaries of
$[0,1]$:
\begin{itemize}
\item $g(0) = 0$,
\item $g^{(i)}(0) = 0$  for all  $i\ge 1$,
\item $g(1) = \int_1^\infty{x^{1-\gamma}}dx = \left.{\frac{1}{2-\gamma}x^{2-\gamma}
}\right|_1^\infty = \frac{1}{\gamma-2}$,

\item $g^{(i)}(1) = \infty$ for all $i\ge 2$.
\end{itemize}
We will approximate $g(t)$ by an
interpolation polynomial $f(t) = \sum_\ell b_\ell t^\ell$.
Obviously, no polynomial has $f^{(i)}(1)=\infty$ for any $i$. Thus we will use
$g$'s values at $t=0,1$, and the derivatives at 0. The
minimal-degree non-trivial polynomial we can use is a cubic
$f(t) = b_0+b_1t+b_2t^2+b_3t^3.$
Since
$g(0) = 0$ we get that $b_0=0$. $g'(0) = 0$ implies $b_1=0$, and
$g''(0) = 0$ implies $b_2=0$.
 Since $g(1) = \frac{1}{\gamma-2}$ we get that $b_3 = \frac{1}{\gamma-2}$. Thus
$f(t) = {t^3}/{(\gamma-2)}$ and
\begin{equation}\label{eq:int}
g(t) = \int_1^\infty{x^{1-\gamma}t^x}dx\approx\frac{t^3}{\gamma-2}. \qed
\end{equation}

\noindent{\bf Notes:}
\begin{itemize}
\item Lemma~\ref{lem:int} approximates the sum of
  Equation~(\ref{eq:basic_sum}) using a cubic polynomial. This is a
  somewhat arbitrary choice: one can use any polynomial of degree
  $d\ge3$ and obtain similar results. Using higher-degree polynomials
  yields a better-quality approximation around $t=0$ since additional
  derivatives are approximated, but does not necessarily improve the
  accuracy of the approximation at $t=1$ since we can only use $g(1)$
  itself. Since we mostly care about the early stages in the
  evolution, near Time $t=1$, we only present the result for the
  special case of $d=3$.
\item At present we give no bound on the error of our polynomial
  approximation.  Instead, to validate our results, in
  Section~\ref{sec:rig} we present a weaker result proven rigorously
  without using the approximation.
\item No polynomial can to give $g^{(i)}(1)=\infty$ for {\em any} $i$, so our
  approximation accuracy is fundamentally limited around $t=1$. It
  would be interesting, and technically more difficult, to
  approximate the sum using rational functions or other functions with an
  asymptote at $t=1$. We leave this to future work.

\end{itemize}

\begin{lem}\label{lem:w}
 Let $\mu$, $a_k$, and $\gamma$ be as in Definition~\ref{def:setup}.
 Then $\mu(\gamma-2)\approx C$.
 \end{lem}
\proof
\begin{eqnarray}\mu(\gamma-2) &=&(\gamma-2)\sum_k{k a_k} =
(\gamma-2)C\sum_k{kk^{-\gamma}}\nonumber\\
 &=& (\gamma-2)C\sum_k{k^{1-\gamma}}  \approx
  (\gamma-2)C\int_1^\infty{x^{1-\gamma}}dx \nonumber\\
 &=&
(\gamma-2)C\left.{\frac{1}{2-\gamma}x^{2-\gamma} }\right|_1^\infty =
(\gamma-2)C\frac{1}{\gamma-2} = C~~.
\end{eqnarray}
(The last equality is valid for $\gamma>2$). $\qed$

In addition to being a building block in proving our main Theorem,
the following Lemma~\ref{lem:pvis} is important since it shows that
most of the edges are detected early during in the sampling process,
near Time $t=1$.

Recall that $P_{vis}(t)$ is the probability that the vertex
discovered at Time $t$ gives rise to {\em another} edge in the BFS
tree---i.e., not the edge it was detected with.
 \begin{lem}\label{lem:pvis}
$P_{vis}(t)\approx t^{3}$
\end{lem}
\proof Recall that $a_k = C\cdot k^{-\gamma}$. Using
equation~(\ref{eq:pvis-approx}) and Lemma~\ref{lem:int} we further
approximate $P_{vis}$.
\begin{eqnarray}
 P_{vis}(t) &\approx&
 \frac{1}{\sum_j{ja_jt^j}}\sum_k{ka_kt^k\left(\frac{\sum_j{ja_jt^j}}{\mu
 t^2}\right)^k}\nonumber\\
 &\approx& \frac{\gamma-2}{C\cdot t^3}\sum_k{C\cdot k^{1-\gamma}\frac{1}{t^k\mu^k}
 \left(\frac{C\cdot t^3}{\gamma-2}\right)^k}
   = \frac{\gamma-2}{t^3}\sum_k{k^{1-\gamma}
    \left(C\cdot \frac{t^{2}}{\mu(\gamma-2)}\right)^k} \nonumber
\end{eqnarray}
and by Lemma~\ref{lem:w} we have that
\begin{eqnarray}
 &\approx& \frac{\gamma-2}{t^3}\sum_k{k^{1-\gamma}t^{2k}}.\label{eq:pvis}
\end{eqnarray}
Let $w=t^{2}$.
By substituting $w$ in equation~(\ref{eq:pvis}) we get
\begin{eqnarray}
P_{vis}(t) &\approx& \frac{\gamma-2}{t^3}\sum_k{k^{1-\gamma}
 w^k}
\end{eqnarray}
Using Lemma~\ref{lem:int} again we get that
\begin{eqnarray}
P_{vis}(t) &\approx& \frac{\gamma-2}{t^3}\frac{w^3}{\gamma-2} =
\left(\frac{w}{t}\right)^3 = t^{3}
\end{eqnarray}
$\qed$

Note that at Time $t=0$ Lemma~\ref{lem:pvis} gives
$P_{vis}\approx 0$, which is as expected: at the end of
the BFS process no new tree-edges are detected. Furthermore, at Time
$t=1$ we get $P_{vis}\approx 1$, again matching our intuition that
at the beginning of the BFS
process the edges detected very likely to be tree edges. Moreover, observe that
most edges are detected at the beginning of the BFS process.

Recall that in the BFS sampling process of \cite{ackm05}
each copy of a vertex $v$ is assigned a Time index $t\in[0,1]$
(Section~\ref{sec:ackmBFS}).
\begin{define}
  Let $\maxindex(v)$ be the maximum index of a vertex $v$, where the maximum is
  taken over all copies of $v$.
\end{define}

\begin{lem}\label{lem:exp-deg}
Let $v$ be a vertex with graph degree $i$ and let $\maxindex(v)=t$. Then
\[
E[deg_T(v) \mid deg_G(v)=i,\maxindex(v)=t]\approx(i-1)t^{3}
\]
\end{lem}
\proof We follow the discussion in~\cite{ackm05}, and neglect the
possibility of self-loops and parallel edges involving a vertex $v$
and its siblings, and ignore the fact that we are choosing without
replacement (i.e., that processing each copy slightly changes the
number of undetected vertices and the number of unmatched copies).
Under these assumptions, the events that each of $v$'s siblings
give rise to edges that will
be detected in the tree are independent, and
\cite{ackm05} shows that the number of
visible edges is approximately binomially distributed as $Bin(i - 1,
P_{vis}(t))$. Therefore
\BEQN
E\left[deg_T(v) \mid deg_G(v)=i,\maxindex(v)=t\right] =
(i-1)P_{vis}(t).\EEQN
Thus, using Lemma~\ref{lem:pvis} it holds
that
\BEQN E\left[deg_T(v) \mid deg_G(v)=i,\maxindex(v)=t\right] \approx
(i-1)t^{3}.
\EEQN $\qed$

\eat{\begin{theo}\label{thm:exp-condition}
Let $m = \frac{1}{2}(i-1)t^{d(d-2)}$, where $d\ge 3$ is a free
parameter. Then for every node $v$ it holds that
\BEQ
\Pr\left[deg_T(v)\ge m \mid deg_G(v)=i,\maxindex(v)=t\right]\ge 1-\eps
\EEQ
where $\eps = e^{-\frac{i-1}{8}t^{d(d-2)}}$
\end{theo}
\proof Recall that the events that each of $v$'s siblings give rise
to a visible edge are approximately independent. Therefore we can
use Chernoff bound. Let $E =
E\left[deg_T(v) \mid deg_G(v)=i,\maxindex(v)=t\right] $. Then \BEQN
\Pr\left[deg_T(v)< m \mid deg_G(v)=i,\maxindex(v)=t\right]&<&
\Pr\left[deg_T(v)<
\frac{1}{2}E \mid deg_G(v)=i,\maxindex(v)=t\right]\\\nonumber
&<&e^{-E/8}\approx e^{-\frac{i-1}{8}t^{d(d-2)}}\EEQN
 $\qed$}

\begin{theo}\label{thm:exp}
Let $v$ be a vertex with graph degree $i$. Then
\BEQ
E\left[deg_T(v) \mid deg_G(v)=i\right]\approx \frac{i(i-1)}{i+3}
\EEQ
\end{theo}
\proof Since $t=\maxindex(v)$ is the maximum of $i$
independent uniform variables in $[0, 1]$, its probability density
is $dt^i/dt = it^{i-1}$. Therefore, using Lemma~\ref{lem:exp-deg},
we get
\BEQN E\left[deg_T(v) \mid deg_G(v)=i\right] \nonumber &=&
\sum_k{k\Pr\left[deg_T(v)=k \mid deg_G(v)=i\right]} \nonumber \\
 &=&
\sum_k{k\int_0^1{it^{i-1}\Pr\left[deg_T(v)=k \mid
      deg_G(v)=i,\maxindex(v)=t\right]}dt} \nonumber \\
&=&
\int_0^1{it^{i-1}\sum_k{k\Pr\left[deg_T(v)=k \mid deg_G(v)=i,\maxindex(v)=t\right]}}dt
\nonumber \\
&=& \int_0^1{it^{i-1}E\left[deg_T(v) \mid
    deg_G(v)=i,\maxindex(v)=t\right]}dt
  \nonumber \\
&\approx& \int_0^1{it^{i-1}(i-1)t^{3}}dt ~~=~~
 i(i-1)\int_0^1{t^{i-1+3}}dt ~~=~~
 \frac{i(i-1)}{i+3}.
\EEQN
This completes the Theorem.
$\qed$

\noindent{\bf Note:} For high-degree nodes $\frac{i}{i+3}\ge \frac{6}{7}$, so
 for high-degree nodes we have that
\BEQ
E\left[deg_T(v) \mid deg_G(v)=i\right]\ge \frac{6}{7}(i-1)
\EEQ
\begin{theo}\label{thm:degT}
Let $m=\frac{i(i-1)}{2(i+3)}$. Then for every node $v$ it holds that
\[\Pr\left[deg_T(v)\ge m \mid deg_G(v)=i\right]\ge 1-\epsilon(i),\]
where $\epsilon(i) = e^{-\frac{i(i-1)}{8(i+3)}}$.
\end{theo}
\proof
 Recall that the events that each of $v$'s copies give rise to a visible edge
are approximately independent. Therefore we can use the Chernoff
bound (cf.\ \cite{mr90}). Let $E = E\left[deg_T(v) \mid deg_G(v)=i\right]$.
Then using Theorem~\ref{thm:exp}, we get
\[
\Pr\left[deg_T(v) < m \mid deg_G(v)=i\right] < \Pr\left[deg_T(v)<
E/2 \mid deg_G(v)=i\right]<e^{-E/8} \approx
e^{-\frac{i(i-1)}{8(i+3)}}.
\] $\qed$

\noindent{\bf Note:} For high-degree nodes ($i\ge 18$)
we have that $\eps(i)\ge 0.16$, and for
$i\ge 32$ we have that $\eps(i)\ge 0.03$.

\eat{As a result of Theorem~\ref{thm:degT} we get that w.h.p the
nodes degrees in the BFS tree received by the sampling process are
close to their original degree in the graph.}

Our main Theorem is now a corollary of Theorem~\ref{thm:degT}.
\eat{\begin{theo} If the underlying graph degree distribution obeys
a power-law then w.h.p the degree distribution of the high-degree
vertices of the sampled graph also follows a power-law, with the
same exponent value.
\end{theo}}

\noindent{\it Proof of Theorem~\ref{thm:power-law}}: As a result of
Theorem~\ref{thm:degT} we get that w.h.p for high-degree nodes
 \[1+\frac{i(i-1)}{2(i+3)}\le deg_T(v)\le i\] where $i = deg_G(v)$.
 Since for high-degree nodes $\frac{i(i-1)}{2(i+3)}\ge
 \frac{6}{7}(i-1)$, we have that w.h.p for high-degree nodes \[1+\frac{3}{7}(i-1)\le deg_T(v)\le i\]
 Therefore w.h.p for high-degree nodes $deg_T(v)\approx 1+c(i-1)$, where $c$ is a constant
 s.t
 $\frac{3}{7}\le c\le 1$. Thus w.h.p for high-degree nodes
\BEQN \Pr\left[deg_T(v)=k\right] \approx
 \Pr\left[deg_G(v)=\frac{k-1}{c}+1\right] =
   \left(\frac{k-1}{c}+1\right)^{-\gamma}\propto k^{-\gamma}.
\EEQN  $\qed$


\section{A More Rigorous Analysis} \label{sec:rig}

Our analysis of the bias, and especially Lemma~\ref{lem:int}, used a
somewhat cavalier polynomial approximation. In this section we give an
alternative derivation of the conservation of the power law tail
behavior without relying on the polynomial approximation of the sum.
We use a more rigorous approach, but we show a weaker result---that
validates the approximations up to multiplicative constants for large
$k$.

\begin{lem}\label{lem:int_new}
If $\gamma>2$ then $Ct\leq C\sum_k{k^{1-\gamma}}t^k\leq \mu,$
for all $0\le t \le 1$.
\end{lem}
\proof
All summands are positive, so the sum is larger than the first summand. Also the
sum is increasing with $t$ and equals $\mu$ for $t=1$.
\QED

\begin{lem}\label{lem:pvis_new}
$P_{vis}(t) \geq\frac{C^2}{\mu^2}$ for all $0\leq t\leq 1$
\end{lem}
\proof Recall that $a_k = C\cdot k^{-\gamma}$. Using
equation~(\ref{eq:pvis-approx}) and Lemma~\ref{lem:int_new} we further
approximate $P_{vis}$ as follows.
\begin{eqnarray}\label{eq:pvis_new}
 P_{vis}(t) &\approx& \frac{1}{\sum_j{ja_jt^j}}
 \sum_k{ka_kt^k\left(\frac{\sum_j{ja_jt^j}}{\mu t^2}\right)^k}\nonumber\\
 &\geq& \frac{1}{\mu}\sum_k{C\cdot k^{1-\gamma}t^k
 \left(\frac{Ct}{\mu t^2}\right)^k} \nonumber\\
 &\geq& \frac{C^2}{\mu^2}.
\end{eqnarray}

\begin{theo}\label{thm:exp_new}
For large enough $k$ there exists $c_1>0$ such that
\[
 c_1 k^{1-\lambda} \leq Pr[deg_T(v)\geq k]\leq Ck^{1-\lambda}
\]
for a random $v$.
\end{theo}
\proof
The upper value follows immediately from the fact that the visible degree is at
most the graph degree.
By Lemma~\ref{lem:pvis_new} we have that $P_{vis}(t)\geq {C^2}/{\mu^2}$
for all $0\leq t\leq 1$. Therefore, for a random $v$ it follows
that
\[
 E[deg_T(v)]\geq \frac{C^2}{\mu^2}deg_G(v)
\]
and hence,
\[
E[deg_G(v)-deg_T(v)]\leq\left(1-\frac{C^2}{\mu^2}\right)deg_G(v).
\]
By the Markov inequality this
means that
\[
Pr[deg_G(v)-deg_T(v)>\alpha]<
 \frac{\left(1-\frac{C^2}{\mu^2}\right)deg_G(v)}{\alpha}.
\]
Take $\alpha=\left(1-\frac{C^2}{\mu^2}+\epsilon\right)deg_G(v)$ for
some constant $\epsilon$. The probability of a node $v$ to have such a
difference between its tree-degree and its graph-degree is at most
some constant less than 1, and therefore, a constant fraction of the
nodes have a degree proportional to the original degree.  Therefore,
the tail of the distribution has a power law with exponent at least
$\gamma$.  $\qed$

Notice that for all $\gamma'<\gamma$, there exists some large $K_*$, such that
$C'K_*^{\gamma'}>CK_*^\gamma$. Therefore, the exponent of the
power law can not decrease throughout the entire degree sequence.

In fact, since the high degree nodes are discovered almost surely at $t\approx
1$, we expect to see $P_{vis}\approx 1$ for these nodes, and therefore, the
behavior of the tail is almost unchanged. Giving an exact bound near $t=1$ is
deferred to a future work.


\begin{figure}[t]
\begin{center}
 \includegraphics[width=4in]{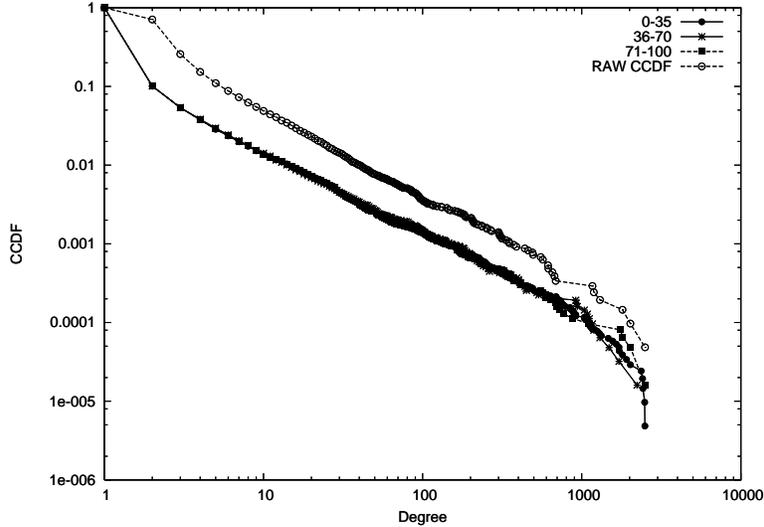}
\end{center}
\caption {CCDF graphs for BFS trees and raw DIMES data}
\label{fig:pdist}
\end{figure}
\begin{table}[t]

\begin{center}
\begin{tabular}{|l|c|}
\hline
\textbf{Data Source} & $\mathbf{\gamma}$ \\
\hline  Group 1 & $ 2.101$ \\
\hline  Group 2 & $ 2.079$ \\
\hline  Group 3 & $ 2.072$ \\
\hline Raw DIMES Data & $ 2.126$ \\
\hline
\end{tabular}

\caption{Sampled power law exponent $\gamma$ for each group.}
\label{tab:gamma}
\end{center}
\end{table}

\section{Simulation Results} \label{sec:Implementation}

To further validate our analysis, we conducted a simulation study.
We used the data collected by Shavitt and Shir \cite{ss05} in the
DIMES project as our underlying graph.
\eat{
paradigm. Each node was examined and assigned a value, where for a
graph with n nodes
$\emph{\textbf{Value}}_{node}=\frac{deg_G(v)}{\sum_n{deg_G(v)}}$.
}

To test whether the choice of the BFS tree root has a noticeable
effect on the resulting degree distribution,
the graph vertices were split into the following 3 groups, based on
their graph degree:
   \begin{enumerate}
    \item Low-degree nodes: $1  \leq   deg_G(v)  <  35$,
    \item Medium-degree nodes: $36  \leq  deg_G(v)  <  70$,
    \item High-degree nodes: $deg_G(v) \ge 71$.
    \end{enumerate}
 From each group we selected 10 nodes at random and constructed a BFS tree
for each, where the selected node was a tree root. An average
CCDF\footnote{Cumulative Complementary Distribution Function.}
of degree distribution was then calculated for every group. We
compared the resulting curves to the
original connectivity data, collected by DIMES.

Figure~\ref{fig:pdist} shows the plotted CCDF curves for the three
groups, and the curve for the raw DIMES data. The figure clearly shows
the familiar power-law curves in all cases, and we can see that the
curves are almost parallel graphs, indicating a similar value of the
power-law exponent $\gamma$.

Table~\ref{tab:gamma} contains the computed values of $\gamma$ for
each group. We can immediately see that the values of $\gamma$ on the
sampled trees (2.072--2.101) are very close to the true power-law
exponent (2.126), thus validating our analysis that the bias is minor.
Furthermore, Table~\ref{tab:gamma} shows that the $\gamma$ values for
the three groups are all close to one another, with a minor decrease
in value as the degree of the root grows. Thus, it seems that the
value of the power-law exponent in the sampled tree is largely
invariant to the degree of the tree root.


\section{Conclusions and Future Work} \label{sec:Conclusions}

We have shown that if the underlying graph degree distribution obeys a
power-law with an exponent $\gamma > 2$ (as is the case in the
AS-graph) then with w.h.p the degree distribution of the high-degree
vertices of the sampled graph also follows a power-law, with the same
exponent value. Therefore, the bias observed in tree-sampling
regarding the degree distribution of the sampled graph is not
significant under these conditions.  Furthermore, since according the
non-tree-sampled data of~\cite{ss05} the AS-graph degree
distribution {\em does} obey a power-law with an exponent $\gamma$ between 2
and 3, we conclude that the bias observed in the degree distribution of the
BGP data is negligible. Thus, the commonly held view of the
Internet's topology as having a degree distribution of a power-law
form with an exponent $2<\gamma<3$ seems to be correct, and unlikely
to be a by-product of the BGP data collection process.

\noindent{\bf Acknowledgment}: we thank Sagy Bar for producing the CCDF graphs
from the DIMES data.

{\small
\bibliographystyle {alpha}
\bibliography{vis}

\newcommand{\etalchar}[1]{$^{#1}$}
\begin{thebibliography}{ACKM05}

\bibitem[AB00]{ab00}
R\'{e}ka Albert and Albert-L\'{a}szl\'{o} Barab\'{a}si.
\newblock Topology of evolving networks: Local events and universality.
\newblock {\em Physical Review Letters.}, 85(24):5234--5237, December 2000.

\bibitem[ACKM05]{ackm05}
D.~Achlioptas, A.~Clauset, D.~Kempe, and C.~Moore.
\newblock On the bias of traceroute sampling.
\newblock In {\em Symposium on the Theory of Computing (STOC)'2005}, Baltimore,
  MD, USA, May 2005.

\bibitem[BA99]{ba99}
Albert-L\'{a}szl\'{o} Barab\'{a}si and R\'{e}ka Albert.
\newblock Emergence of scaling in random networks.
\newblock {\em {Science}}, 286:509--512, 15 October 1999.

\bibitem[BB01]{bb01}
G.~Bianconi and A.~L. Barab\'{a}si.
\newblock Competition and multiscaling in evolving networks.
\newblock {\em Europhysics Letters}, 54(4):436--442, 2001.

\bibitem[BBBC01]{bbbc01}
P.~Barford, A.~Bestavros, J.~Byers, and M.~Crovella.
\newblock On the marginal utility of network topology measurements.
\newblock In {\em Proc. ACM SIGCOMM}, 2001.

\bibitem[BGW04]{bgw04}
S.~Bar, M.~Gonen, and A.~Wool.
\newblock An incremental super-linear preferential {Internet} topology model.
\newblock In {\em Proc.\ 5th Annual Passive \& Active Measurement Workshop
  (PAM), LNCS 3015}, pages 53--62, Antibes Juan-les-Pins, France, April 2004.
  Springer-Verlag.

\bibitem[BGW05]{bgw05b}
S.~Bar, M.~Gonen, and A.~Wool.
\newblock A geographic directed preferential {Internet} topology model.
\newblock In {\em Proc.\ 13th IEEE Symp. Modeling, Analysis, and Simulation of
  Computer and Telecommunication Systems (MASCOTS)}, pages 325--328, Atlanta,
  GA, September 2005.

\bibitem[Bol85]{b85}
B.~Bollob\'{a}s.
\newblock {\em Random Graphs}.
\newblock Academic Press, 1985.

\bibitem[BS02]{bs02}
R.X. Brunet and I.M. Sokolov.
\newblock Evolving networks with disadvantaged long-range connections.
\newblock {\em Physical Review E}, 66(026118), 2002.

\bibitem[BSH03]{CIDR}
Tony Bates, Philip Smith, and Geoff Huston.
\newblock Cidr reports, 2003.
\newblock \url{http://www.cidr-report.org/}.

\bibitem[BT02]{bt02}
T.~Bu and D.~Towsley.
\newblock On distinguishing between {Internet} power-law generators.
\newblock In {\em Proc. IEEE INFOCOM'02}, New-York, NY, USA, April 2002.

\bibitem[CCG{\etalchar{+}}02]{ccgjsw02}
Q.~Chen, H.~Chang, R.~Govindan, S.~Jamin, S.~Shenker, and W.~Willinger.
\newblock The origin of power laws in {Internet} topologies revisited.
\newblock In {\em Proc. IEEE INFOCOM'02}, New-York, NY, USA, April 2002.

\bibitem[CM04a]{cm04}
A.~Clauset and C.~Moore.
\newblock Traceroute sampling makes random graphs appear to have power law
  degree distributions.
\newblock 2004.
\newblock Preprint, Submitted to Physical Review Letters.

\bibitem[CM04b]{cm04b}
A.~Clauset and C.~Moore.
\newblock Why mapping the {Internet} is hard, 2004.
\newblock arXiv:cond-mat/0407339.

\bibitem[ER60]{er60}
P.~Erd\H{o}s and A.~Renyi.
\newblock On the evolution of random graphs.
\newblock {\em Magyar Tud.\ Akad.\ Mat.\ Kutato Int.\ Kozl.}, 5:17--61, 1960.

\bibitem[FFF99]{fff99}
C.~Faloutsos, M.~Faloutsos, and P.~Faloutsos.
\newblock On power-law relationships of the {Internet} topology.
\newblock In {\em Proc. of ACM SIGCOMM'99}, pages 251--260, August 1999.

\bibitem[GT00]{gt00}
R.~Govindan and H.~Tangmunarunki.
\newblock Heuristics for {Internet} map discovery.
\newblock In {\em Proc. IEEE INFOCOM'00}, pages 1371--1380, Tel-Aviv, Israel,
  March 2000.

\bibitem[KRR01]{krr01}
P.L. Krapivsky, G.J. Rodgers, and S.~Render.
\newblock Degree distributions of growing networks.
\newblock {\em Physical Review Letters}, 86:5401, 2001.

\bibitem[LBCX03]{lbcx03}
A.~Lakhina, J.~W. Byers, M.~Crovella, and P.~Xie.
\newblock Sampling biases in {IP} topology measurments.
\newblock In {\em Proc. IEEE INFOCOM'03}, 2003.

\bibitem[LC03]{lc03}
X.~Li and G.~Chen.
\newblock A local-world evolving network model.
\newblock {\em Physica A}, 328:274--286, 2003.

\bibitem[MR90]{mr90}
R.~Motwani and P.~Raghavan.
\newblock {\em Randomized Algorithms}.
\newblock Cambridge University Press, 1990.

\bibitem[PR04]{pr04}
T.~Petermann and P.~De~Los Rios.
\newblock Exploration of scale-free networks - do we measure the real
  exponents?
\newblock {\em Europhysics Letters}, 38:201--204, 2004.

\bibitem[RN04]{rn04}
H.~Reittu and I.~Norros.
\newblock On the power law random graph model of the {Internet}.
\newblock {\em Performance Evaluation}, 55, January 2004.

\bibitem[Rou05]{routeViews}
University of {Oregon} {Route Views} project, 2005.
\newblock \url{http://www.routeviews.org}.

\bibitem[SS05]{ss05}
Y.~Shavitt and E.~Shir.
\newblock {DIMES}: Let the {Internet} measure itself, 2005.
\newblock \url{http://www.arxiv.org/abs/cs.NI/0506099}.

\bibitem[TGJ{\etalchar{+}}02]{tgjsw02}
H.~Tangmunarunkit, R.~Govindan, S.~Jamin, S.~Shenker, and W.~Willinger.
\newblock Network topology generators: Degree based vs.\ structural.
\newblock In {\em Proc. ACM SIGCOMM}, 2002.

\bibitem[TPSF01]{tpsf01}
L.~Tauro, C.~Palmer, G.~Siganos, and M.~Faloutsos.
\newblock A simple conceptual model for {Internet} topology.
\newblock In {\em IEEE Global Internet}, San Antonio, TX, November 2001.

\bibitem[WGJ{\etalchar{+}}02]{wgjps02}
W.~Willinger, R.~Govindan, S.~Jamin, V.~Paxson, and S.~Shenker.
\newblock Scaling phenomena in the {Internet}: Critically examining
  criticality.
\newblock {\em Proceedings of the National Academy of Sciences of the United
  States of America}, 99:2573--2580, February 2002.

\bibitem[WJ02]{wj02}
Jared Winick and Sugih Jamin.
\newblock Inet-3.0: {Internet} topology generator.
\newblock Technical Report UM-CSE-TR-456-02, Department of EECS, University of
  Michigan, 2002.

\end{thebibliography}
}

\end{document}